\documentstyle[psfig]{article}

\begin{document}

\def\gsim{\!\!\!\phantom{\ge}\smash{\buildrel{}\over
  {\lower2.5dd\hbox{$\buildrel{\lower2dd\hbox{$\displaystyle>$}}\over
                               \sim$}}}\,\,}
\def\kms{\rm ~km~s^{-1}}
\def\Mdot{\dot M}

\begin{center}

\Large
{\bf A cosmological surprise: the universe accelerates.}
\normalsize

\vspace{1.0cm}
Bruno Leibundgut \& Jesper~Sollerman\\

\end{center}


\noindent{European Southern Observatory, Karl-Schwarzschild-Strasse 2,
D-85748 Garching bei M\"unchen, Germany}\\

\vspace{1.0cm}


Cosmology is in turmoil. The standard model of a few years ago has
recently been abandoned and replaced by new ideas. The reasons for
this dramatic change are new measurements of the geometry and the
matter contents of the universe. The new model implies a dynamical age
of the universe that accommodates the oldest known stellar objects,
but raises the need for a dark energy component, which is not readily
explained within the current particle physics theories.

Most current cosmological models are based on the Big Bang theory, in
which the universe started in a hot and dense state. Since then, the
cosmic expansion has led to adiabatic cooling, and the subsequent
condensations of the matter have formed stars, galaxies and clusters
of galaxies.  The Big Bang theory predicts not only the universal
expansion, but also the baryonic matter content and a relic radiation
from the original hot phase. All of these predictions have been
observed. The detection of the cosmic microwave radiation and the tiny
temperature fluctuations in it are often cited as the most spectacular
success of this theory (e.g. Peebles 2001). An extension to the Big
Bang theory is the proposed inflationary phase at the very earliest
times of the expansion. The inflation would be driven by energy, which
emerges from the decay of a particle field and produces a universe
many times larger than a simple linear expansion. Inflation predicts
the seeds for the growth of large-scale structure in the
universe. Since the universe would be inflated several orders of
magnitudes, it would essentially have a flat space structure (Guth
1997). Among the open questions in this picture are the matter/energy
content of this universe and its future.

The average energy density determines the fate of a homogeneous and
isotropic universe governed by gravitational forces. It was thought
that the matter content governs the expansion of the universe today,
while at early phases radiation dominated. In this simple picture the
geometry of the universe is directly coupled to the matter density. A
flat universe implies a density of about 
$8\times10^{-27}$~kg~m$^{-3}$,
the critical
density. For an average density below this value, the universe would
be open and expand forever, while at a higher density it would be
closed and eventually re-collapse.

Just a few years ago the total energy content of the universe was
unknown. All indications were that baryonic matter makes up only about
5$\%$ of the critical density (Bahcall et al. 1999). To explain the
kinematics of galaxies and clusters of galaxies an additional mass
component often referred to as "missing" or "dark" mass had to be
introduced. The search for direct evidence for the dark matter has,
despite substantial efforts, been futile so far. Nevertheless, from
observations of the evolution and mass determinations of clusters of
galaxies, it appears that dark and baryonic matter could explain about
30$\%$ of the critical density. Should the universe be flat, as required
if inflation is correct, then some other energy component must
contribute to the average density.

For a cosmologist, there are only few parameters needed to describe
the universe. All models are based on Einstein's theory of general
relativity. The world models are characterized by two parameters: the
current rate and the deceleration of the expansion. The first
parameter is called the Hubble constant after Edwin Hubble, who
discovered the cosmic expansion in 1929. The other parameter describes
the change of the expansion and depends on the energy density and the
curvature of the universe. The contributions to the density are
expressed as fractions of the critical density and are denoted by the
Greek letter $\Omega$, e.g. $\Omega_{\rm M}$, for the matter density. The expansion itself
is typically measured by the redshift. This is the ratio of the scale
factor at two different times of the expansion and observationally a
shift of spectral features to longer wavelengths. Hubble's law states
that for small distances the redshift is proportional to the distance.

At large look-back times and distances the linearity of Hubble's law
breaks down and the distances depend on the energy density of the
universe. The various constituents, typically matter and radiation are
considered, contribute in different ways to the energy
density. Radiation ceased to be gravitationally important at a
redshift of about 1000, a time from which we can only measure the
cosmic microwave background radiation. Another component is the famous
cosmological constant introduced by Albert Einstein to reconcile the
solutions of his equations with a static universe. He later abandoned
this term, when Edwin Hubble discovered the general expansion of the
universe. For many decades the cosmological constant was not
considered in the world models as there was not obvious reason to
include it and as it was not possible to connect it to any particle
theory. In modern terms, it represents the contribution of the vacuum
energy (Carroll et al. 1992).

The last three years have seen truly exciting progress in
observational cosmology. The flat geometry of the universe has been
confirmed by balloon-borne experiments that measured the fluctuations
in the cosmic microwave background. The physical scale of these
fluctuations can be determined from first principles and the
measurement of their angular extend on the sky gives a direct
indication of the geometry of the universe. All experiments agree that
the universe is most likely flat (de Bernardis et al. 2000, Balbi et
al. 2000).

A complementary approach to determine the geometry is through the
measurement of distances. This approach actually measures the
deceleration due to the gravitational attraction, which slows down the
cosmic expansion. The amount of deceleration hence directly determines
the average density.

Cosmological distances are, however, notoriously difficult to
attain. The long and arduous history of the determination of the
Hubble constant painfully reflects this difficulty. Even today, it
appears that the value of the Hubble constant is uncertain to about
10$\%$. The easiest way to measure distances is through standard candles,
i.e. objects with identical absolute luminosity. Many candidates for
standard candles have been proposed. Only few astronomical objects
have turned out to be suitable.

A prime candidate for such a standard candle is a certain type of
stellar explosions. Known to astronomers as Type Ia Supernovae these
explosions occur at the end stages of stellar evolution, when low-mass
stars exhaust their fuel and start to contract and cool down. The
resulting white dwarfs are so compact and dense that they are
supported by electron pressure. It can be shown that there is an upper
limit to the mass supported by the pressure of the degenerate
electrons. This Chandrasekhar limit, named after Subrahmanyan
Chandrasekhar, is near 1.4 times the mass of the Sun. A single,
isolated white dwarf will not change its mass, but a white dwarf in a
double star system may acquire mass from its companion. If this
process is efficient enough, the white dwarf will turn itself into a
thermonuclear bomb that will burn carbon and oxygen explosively to
nuclear statistical equilibrium. For the density and pressure in these
explosions this is mostly a radioactive nickel isotope ($^{56}$Ni),
which decays through gamma-decay to $^{56}$Co and further to stable 
$^{56}$Fe. The
energy is deposited in the star and blows it apart. Such an explosion
can outshine the light of a whole galaxy made up of some 10$^{10}$ stars
(Figure 1). The Chandrasekhar mass is a natural limit and makes it
conceivable that all Type Ia Supernovae are similar. Such a uniform
constellation assures that there are only small differences between
individual explosions, a prime condition for a standard candle.

\section{Observing supernovae}

Supernovae are extremely rare. A galaxy like our Milky Way may produce
a Type Ia Supernova only every 400 years. Hence, one has to observe a
large number of galaxies for quite some time to detect a
supernova. The sample of nearby supernovae (out to a redshift of about
z=0.1) is still rather limited. In fact, well-sampled light curves
(e.g. Figure 1) are an exception and it has taken over a decade of
dedicated effort to collect a significant sample of objects. In the
process a lot has been learnt about Type Ia Supernovae and their
physics (for a review see Leibundgut 2000).

Not all Type Ia Supernovae are identical. However, there is a way to
normalise their peak luminosity according to their light curves. More
luminous supernovae display a slower temporal evolution. In this way,
it is possible to normalise all objects to the same peak luminosity
and make them exquisite standard candles.

One way to empirically test the quality of a standard candle is to
look at the distance vs. redshift diagram (Figure 2).  For a redshift
below z=0.1 the scatter around the line of linear expansion is
extremely small and proves that, at least in the nearby universe, the
supernovae can be used as distance indicators. Direct determination of
the absolute luminosity of Type Ia Supernovae has recently been
obtained for the 10 nearest supernovae which exploded within the last
60 years. The peak luminosity of these objects is indeed very uniform.

Distant supernovae, like the example in Figure 1, are more difficult
to observe. Since supernovae are such rare events a large volume has
to be surveyed to discover a sufficient number for co-ordinated
follow-up observations. It was soon recognised that such a project
would exceed the possibilities of a single observatory or group of
astronomers. In fact, this experiment makes use of all the largest
telescopes available. A typical campaign involves the European
Southern Observatory VLT (Figure 3), the Keck telescopes, the
Canada-France-Hawaii telescope, the Cerro Tololo Inter-American
Observatory and the Hubble Space Telescope. The supernovae are
observed for about two months with all available telescopes. The
resulting light curves, like the one shown in Figure 1, are analysed
to derive the peak brightness and plot them in diagrams like Figure 2.

With the model lines calibrated by the nearby supernovae the distant
explosions tell us how much the universe has expanded between then and
now. From Figure 2 it is obvious that the distances are larger than
what would be expected in a freely coasting, i.e. empty,
universe. Even worse would be a fit for a flat and matter-filled
Einstein-de Sitter model. The distant supernovae are simply too faint,
i.e. too distant to be compatible with the old paradigm of a universe
filled with matter and radiation only. There has to be a component to
the energy density that has accelerated the expansion over the last
$\sim6\times10^{9}$ years (Riess et al. 1998, Perlmutter et al. 1999, Hogan et
al. 1999).

An obvious candidate is the cosmological constant. With the supernova
result it appears that it should be re-introduced to explain the
data. Another possibility is that the cosmological constant is indeed
zero, but that there is a particle field that through its decay acts
like a cosmological constant. It has now come to be known as
'quintessence' (Ostriker and Steinhardt 2001). Independent of which
explanation is correct we can designate the energy density of this
``dark energy'' as $\Omega_{\Lambda}$. 
Figure 4 shows the probability distribution
between the matter density $\Omega_{\rm M}$  and $\Omega_{\Lambda}$ as the 
supernova distances define
them. It is obvious that models with  $\Omega_{\Lambda}$=0 are excluded at 
the $>95\%$
level for all models with a positive matter content. This comes from
the fact that an accelerating, negative pressure component is
required, which can not be achieved with matter or radiation.

A nice feature of world models with these parameters would be that the
dynamical age of the universe is no longer in conflict with the ages
of the oldest stars in our Milky Way. We have marked the isochrones of
the dynamical age for the combination of $\Omega_{\rm M}$  and $\Omega_{\Lambda}$.

The result from the distant supernovae that the expansion of the
universe is not decelerating but in fact accelerates is certainly
surprising. We should answer the obvious question how secure is this
result is. There are a few other possible explanations for the
faintness of the distant supernovae. They are an intrinsic evolution
of the peak brightness of the supernovae, unrecognised dust, or
gravitational lensing. Evolution has been the downfall of all
previously proposed distance indicators. Take for example regular
galaxies. Looking back for several 109 years means that a galaxy is
composed of stars, which are on average younger than the ones
currently in our Milky Way. Since a galaxy with young stars has a
larger fraction of short-lived, massive stars, which are brighter and
bluer, it will change its luminosity over time. Could Type Ia
Supernovae also suffer from evolutionary effects? At first glance,
such an evolution is less likely. According to the current models
they are explosions of stars, which always end up in the same
configuration, a white dwarf at the Chandrasekhar limit. However,
even though the bomb always has the same mass, its composition may
vary slightly. The explosive carbon burning in a white dwarf depends
on the mixture of carbon and oxygen. One could imagine that the
distant supernovae have had a different chemical composition at the
explosion than their nearby counterparts. Also, stellar evolution
predicts that younger white dwarfs are the descendants from
predominantly more massive stars. Thus, the parent population of the
distant supernovae could be different from the one observed
nearby. Unlike for the nearby supernovae we can not use the linear
expansion law to check for consistency. What has been tried in the
last few years is to make sure that the distant objects have the same
appearance as the nearby ones. This program is still under way, but
first results are already becoming available. The observed spectra of
the two populations are so far indistinguishable (Figure 5). 
There is further a very characteristic light curve shape with a
second maximum of Type Ia Supernovae in the near-infrared, which by
now has also been observed for the distant supernovae. On the other
hand, there appears to be a trend to bluer colours for distant
supernovae, which, if confirmed, may hint at evolution.

Interstellar dust also dims astronomical objects. Galactic dust
distributed in the plane of the Milky Way is the reason we can not
observe the Galactic Centre in optical light. But galactic dust not
only dims the objects, it also preferably scatters blue light and
makes objects appear redder. If the intrinsic colour of an object is
know, the observed light can be corrected according to the galactic
absorption law. The colour evolution of Type Ia Supernovae is in fact
very uniform and by always obtaining colour information for the
distant supernovae we can check for dust and correct the brightness,
if necessary. But there is nothing that says that the galactic dust is
representative throughout the universe. For example, larger dust
particles scatter differently so that the supernovae still would be
dimmed but not as reddened. Observing programs to check this
possibility have been carried out last year and are currently being
analysed.

Along the light path of any distant object lie other massive
objects. Gravitational lensing of distant objects has been observed in
several forms. Background objects are distorted by the potential wells
through which light has to travel. Large arcs of distorted background
galaxies are now regularly observed in massive clusters of
galaxies. Another effect of lensing is amplification or
de-amplification of the light. It turns out that distant objects are
on average de-amplified, i.e. statistically appear fainter. This would
be an obvious explanation of the dimness of the distant
supernovae. However, the effect is not large enough, even if all mass
would be concentrated in compact objects, and can not explain the
distant supernovae. This result is based on model calculations, as
this is the only effect that can not be observed with the supernova
light itself.

There is one signature, which would almost unambiguously prove that
the universe has been accelerating over about half its age. This is
the deceleration during the early phases of the expansion. The
cosmological constant does not change over time, but since the density
decreases the gravitational action should have been much larger during
the first few 10$^{9}$ years. In Figure 2 this would show up as a
non-monotonic evolution. The critical redshift range is around z=1.2,
a range that has so far not been explored by the observations. This is
stretching the current capabilities of any existing telescope. The
supernovae become exceedingly faint and due to the redshift have to be
observed at near-infrared wavelengths where the night sky is much
brighter than in the optical. Nevertheless, the next observing
campaigns will target exactly this redshift range to test whether the
acceleration of the universal expansion is indeed the correct
interpretation for the dimness of the distant supernovae.

There is a big problem with the cosmological constant. In modern
particle theories it is associated with the energy of the vacuum, but
these theories also predict a value for the vacuum energy which
deviates by more than 50 orders of magnitudes from the cosmological
observations. Although this has been a problem all along, the
supernova measurements have acerbated this problem, by requiring a
non-zero but small ($<1$) cosmological constant. For these reasons many
theorists currently favour quintessence models. But this requires
introducing a new particle and corresponding potentials, which have to
be fine-tuned to have an action like the observed one.

But there is more on the horizon. To distinguish between a
cosmological constant and quintessence the time variability of the
acceleration should be checked. Tracing the supernova distances in
detail can do this. A large sample of supernovae out to a redshift of
about z=1.5 is needed. Proposals to obtain such samples have already
been made.

Type Ia Supernovae together with the recent results on the cosmic
microwave background and the masses of clusters of galaxies are
consistent with a flat universe, in which about 30$\%$ is gravitating
matter and 70$\%$ is contributed by ''dark energy'' (cosmological
constant or quintessence). Only about 5$\%$ of the total energy stem from
baryonic matter. Another 5$\%$ may be contributed by massive neutrinos.

The opinions of cosmologists currently range from visions of
''precision cosmology'' to worries about the fact that we have to add
new constituents to the universe for which we have currently no
explanation at all. This is not necessarily a contradiction. Observers
have been furnished with tools over the last decade, which allow them
to probe many of the cosmological questions in much more detail and
with much higher precision. On the other hand, these new results have
shown that our picture of the universe was incomplete and will need
further scrutiny.

\clearpage

\begin{figure}[h]
\centering
\psfig{file=fig1.ps,%
bbllx=0mm,bblly=0mm,bburx=540mm,bbury=270mm,width=300mm}
\vspace{-2.0 cm}
\caption{ A series of observations of a distant Type Ia Supernova. The
object is located in a galaxy at a redshift of z=0.51 corresponding to
a look back time of about 40$\%$ the age of the universe. The
observations were obtained with the European Southern Observatory 3.5m
diameter New Technology Telescope at the La Silla Observatory in
Chile. The change in image sharpness is caused by the varying
atmospheric conditions in the individual nights. The supernova at its
brightest phase is about 15 million times fainter than what can be
seen by naked eye.  The light curve of this object has been derived
from the images. This diagram displays the rise and fall of the
supernova brightness as a function of time. The red line indicates the
evolution of a typical nearby Type Ia Supernova.
}
\label{lc.ps}
\end{figure}

\clearpage

\begin{figure}[h]
\centering
\psfig{file=fig2.ps,%
bbllx=0mm,bblly=0mm,bburx=540mm,bbury=270mm,width=230mm}
\vspace{-1. cm}
\caption{ Diagram displaying distance vs. redshift for Type Ia
Supernovae, generally referred to as Hubble
diagrams, as it is similar, but not identical to the original
diagram showing the expansion of the universe. The upper panel shows
the conventional form of this diagram. The points are the supernova
measurements assembled by two independent groups, the High-z Supernova
Search Team (Riess et al. 1998) and the Supernova Cosmology Project
(Perlmutter et al. 1999). The nearby supernovae have been measured
over about one decade, while the distant objects have been observed in
short campaigns of a few months length. The lines represent the
expectations for a standard candle in the specific world models. They
represent an Einstein-de Sitter model in which the universe is
completely dominated by matter $\Omega_{\rm M}$=1.0, $\Omega_{\Lambda}$=0.0), an empty universe
$\Omega_{\rm M}$=0.0, $\Omega_{\Lambda}$=0.0) with nothing in it, and a universe dominated by the
cosmological constant $\Omega_{\rm M}$=0.0, $\Omega_{\Lambda}$=1.0). The location of the 
model
lines is set by the data for the nearby supernovae. The dotted line is
a model which is consistent with all current data and has $\Omega_{\rm M}$=0.3 and
$\Omega_{\Lambda}$=0.7.  The lower panels show the same diagram normalised to the
model of an empty universe. It is obvious that the distant supernovae
lie above the line for an empty universe and require some contribution
by the cosmological constant or quintessence. The favoured model
predicts a non-monotonic behaviour of standard candles in this
diagram.}
\label{hzplot.ps}
\end{figure}

\clearpage

\begin{figure}[h]
\centering
\psfig{file=fig3.ps,%
bbllx=0mm,bblly=0mm,bburx=540mm,bbury=270mm,width=300mm}
\vspace{-1.5 cm}
\caption{
The ESO Very Large Telescope (VLT) on Cerro Paranal in Chile. 
This new array of four 8m telescopes is an example of what is required to 
observe the distant supernovae in detail. Only large telescopes, like the 
VLT, the 10m Keck telescopes on Mauna Kea in Hawaii and the Gemini 8m 
telescopes in Hawaii and Chile are capable to gather enough light for a 
detailed analysis. 
}
\label{telescope.ps}
\end{figure}

\clearpage

\begin{figure}[h]
\centering
\psfig{file=fig4.ps,%
bbllx=0mm,bblly=0mm,bburx=540mm,bbury=270mm,width=300mm}
\vspace{-1.0 cm}
\caption{ Probability distribution of $\Omega_{\Lambda}$ vs. $\Omega_{\rm M}$ 
from the supernova
data. The greyscale gives the probability distribution as derived from
the supernovae shown in Figure 2. The contours are drawn at 68.7$\%$,
95.4$\%$ and 99.7$\%$ confidence level. The preference for a ''dark energy''
component is obvious. $\Omega_{\Lambda}$ = 0 is excluded at the $>95\%$ level. The line
for a flat universe as favoured by the cosmic microwave background is
shown in red. The sloped lines indicate the age of the universe for
these models assuming a Hubble constant of H$_{0}$=65~km~s$^{-1}$~Mpc$^{-1}$.
The oldest stars have an age of about 13(109 years, which is consistent
with the new supernova data but not with the Einstein-de Sitter model
$\Omega_{\rm M}$=1, $\Omega_{\Lambda}$=0).}
\label{prob.ps}
\end{figure}

\clearpage

\begin{figure}[h]
\centering
\psfig{file=fig5.ps,%
bbllx=0mm,bblly=0mm,bburx=540mm,bbury=270mm,width=300mm}
\vspace{-1.0 cm}
\caption{Comparison of spectra of nearby and distant
supernovae. Spectra of seven distant Type Ia Supernovae observed with
the VLT (blue) are compared with nearby objects (red) at the same
phase. The distant objects look remarkably similar to the nearby
ones. This is an important check as the spectra are quasi ''finger
prints'' for the chemical composition of the supernova
ejecta. Contamination of galaxy light or reddening of the objects due
to dust can cause changes in the slopes of the continuum emission.}
\end{figure}

\end{document}